\def   \ni {\noindent}

\def   \ssk {\vskip  5truept}

\def   \bsk {\vskip 15truept}

\def   \newline {\hfil\break}

\documentstyle[epsfig]{article}
\begin{document}

\hsize 5truein
\vsize 8truein
\font\abstract=cmr8
\font\keywords=cmr8
\font\caption=cmr8
\font\references=cmr8
\font\text=cmr10
\font\affiliation=cmssi10
\font\author=cmss10
\font\mc=cmss8
\font\title=cmssbx10 scaled\magstep2
\font\alcit=cmti7 scaled\magstephalf
\font\alcin=cmr6 
\font\ita=cmti8
\font\mma=cmr8
\def\ref{\par\noindent\hangindent 15pt}
\null

\setlength{\unitlength}{1mm}
\def\fwb{65mm}
\def\fhb{45mm}
\newcommand{\Dunits}{$\times10^{28}$ cm$^2$s$^{-1}$}
\newcommand{\gray}{$\gamma$-ray\ }
\hyphenation{brems-strahl-ung}

{\footnotesize \it \vspace{-14\baselineskip} \noindent 
   Proc.\ 3rd INTEGRAL Workshop ``The Extreme Universe'',
   14--18 Sep.\ 1998, Taormina, Italy
   \\ \rule[3ex]{124mm}{0.1mm}
\vspace{1ex} \vspace{11\baselineskip} }


\title{\ni PUZZLES OF GALACTIC CONTINUUM {\huge $\gamma$}-RAYS}

\bsk \bsk
\author{\ni I.V. Moskalenko$^{1,2}$, A.W. Strong$^1$}                                                       
\bsk
\affiliation{1) Max-Planck-Institut f\"ur extraterrestrische Physik, 
D-85740 Garching, Germany}                                                

\affiliation{2) Institute for Nuclear Physics, Moscow State University, 
119 899 Moscow, Russia}                                                
\bsk
\baselineskip = 12pt

\abstract{ABSTRACT \ni
Inverse Compton scattering appears to play a more important r\^ole in the
diffuse Galactic continuum emission than previously thought, from MeV
to GeV energies. We compare models having a large inverse Compton
component with EGRET data, and find good agreement in the longitude and
latitude distributions at low and high energies. We test an alternative
explanation for the $\ge$1 GeV \gray excess, the hard nucleon spectrum,
using secondary antiprotons and positrons.  At lower energies to fit
the COMPTEL and OSSE data as diffuse emission requires either a steep
upturn in the electron spectrum below 200 MeV or a population of
discrete sources.
}                                                    
\bsk
\baselineskip = 12pt

\bsk
\baselineskip = 12pt


\text{\ni 1. INTRODUCTION
\ssk
\ni     

We are developing a model which aims to reproduce self-consistently
observational data of many kinds related to cosmic-ray origin and
propagation: direct measurements of nuclei, antiprotons, electrons and
positrons, $\gamma$-rays, and synchrotron radiation (SM98).

Recent results from both COMPTEL and EGRET indicate that inverse
Compton (IC) scattering is a more important contributor to the diffuse
emission that previously believed.  COMPTEL results (Strong et
al.\ 1997) for the 1--30 MeV range show a latitude distribution in the
inner Galaxy which is broader than that of HI and H$_2$, so that
bremsstrahlung of electrons on the gas does not appear adequate and a
more extended component such as IC is required.  The broad distribution
is the result of the large $z$-extent of the interstellar radiation
field (ISRF) which can interact cosmic-ray electrons up to several kpc
from the plane.  At much higher energies, the puzzling excess in the
EGRET data above 1 GeV relative to that expected for $\pi^0$-decay has
been suggested to orginate in IC scattering (e.g., PE98) from a hard
interstellar electron spectrum.

\begin{figure}[!t]
   \begin{picture}(120,45)(5,-45)
      \put(-1,0){ \makebox(60,0)[tl]{ \psfig{file=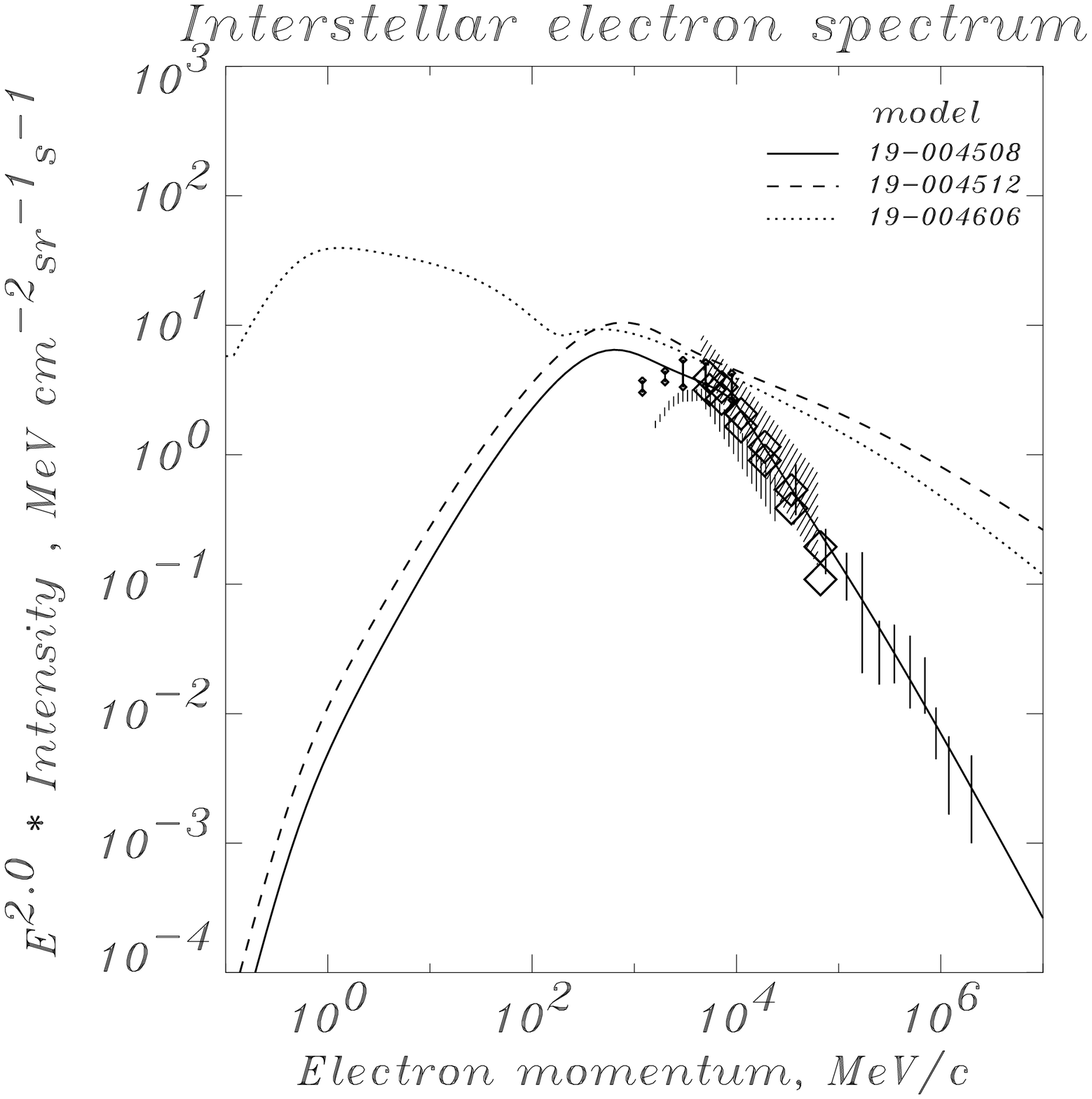,%
         height=\fhb,width=\fwb,clip=}}}
      \put(61,0){ \makebox(60,0)[tl]{ \psfig{file=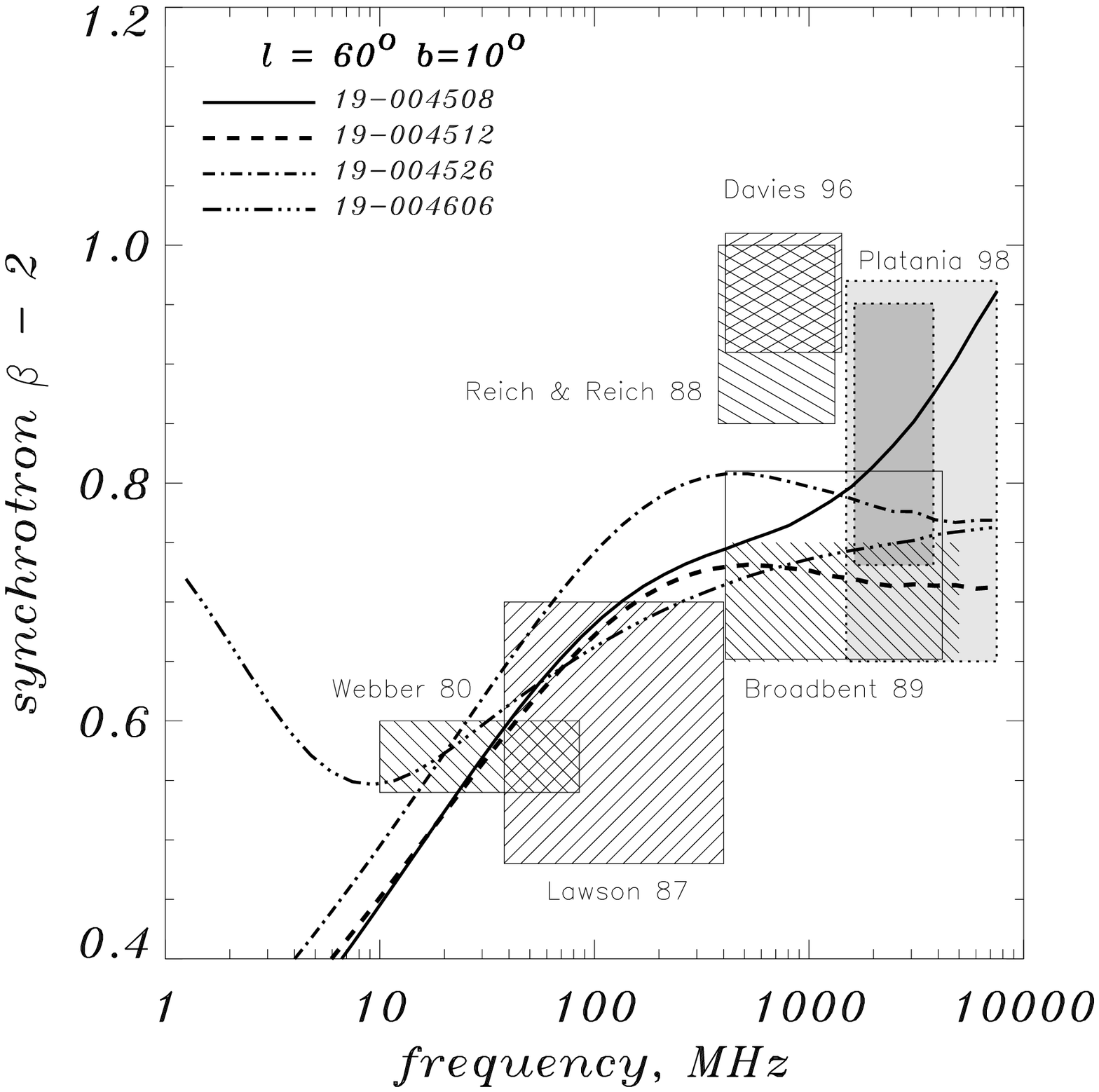,%
         height=\fhb,width=\fwb,clip=}}}
   \end{picture}
\caption{FIGURE 1. 
Left: Electron spectra at $R$=8.5 kpc in the plane,
for `normal' and hard electron spectrum models
with and without low-energy upturn.
Data points:  direct measurements, see references in MS98.
Right: Synchrotron spectral index for a representative direction for
these electron spectra, compared to data.
\label{fig1} }
\end{figure}
\begin{figure}[t!]
   \begin{picture}(120,45)(5,-45)
      \put(-1,0){ \makebox(60,0)[tl]{ \psfig{file=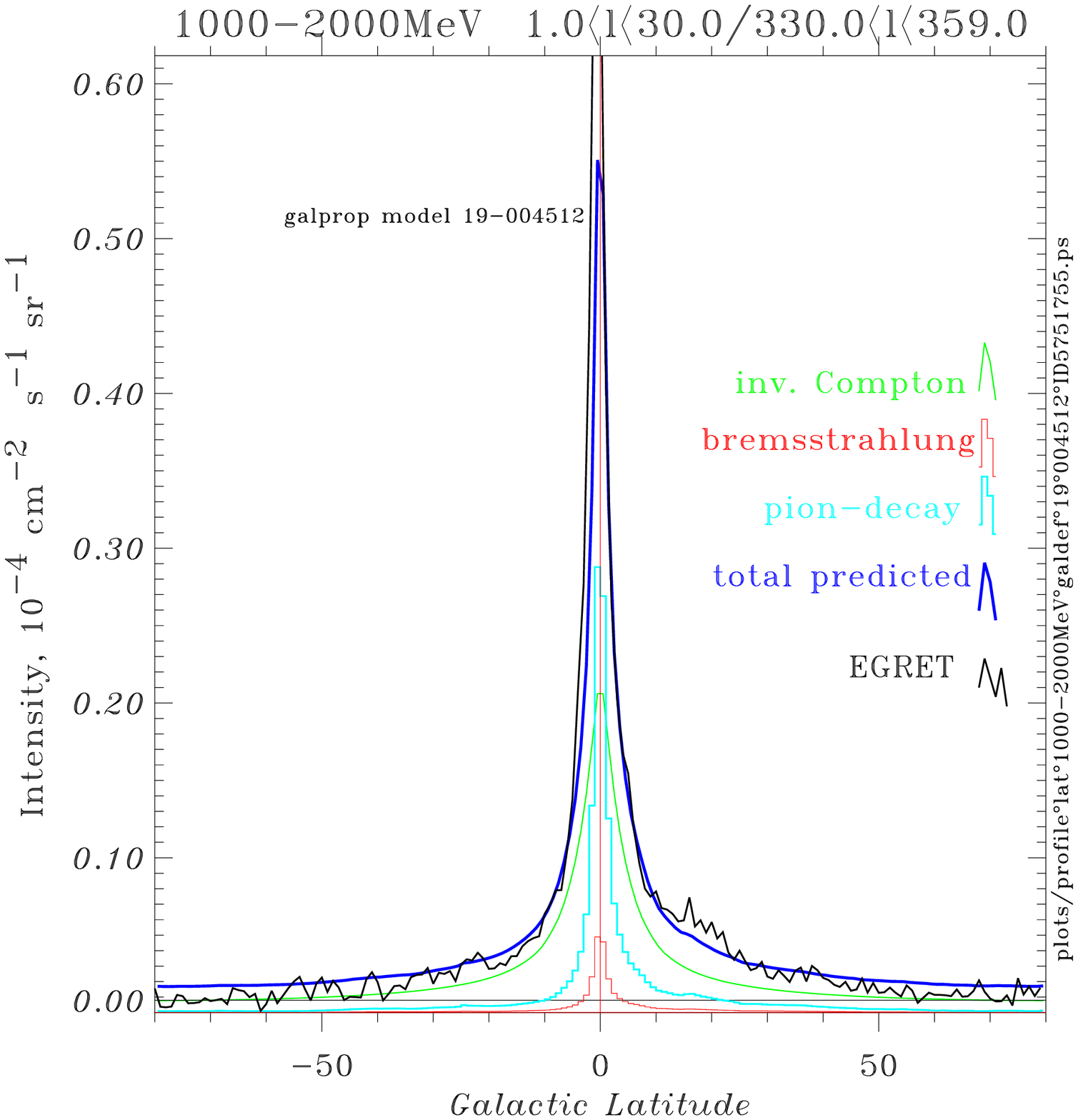,%
         height=\fhb,width=\fwb,clip=}}}
      \put(61,0){ \makebox(60,0)[tl]{ \psfig{file=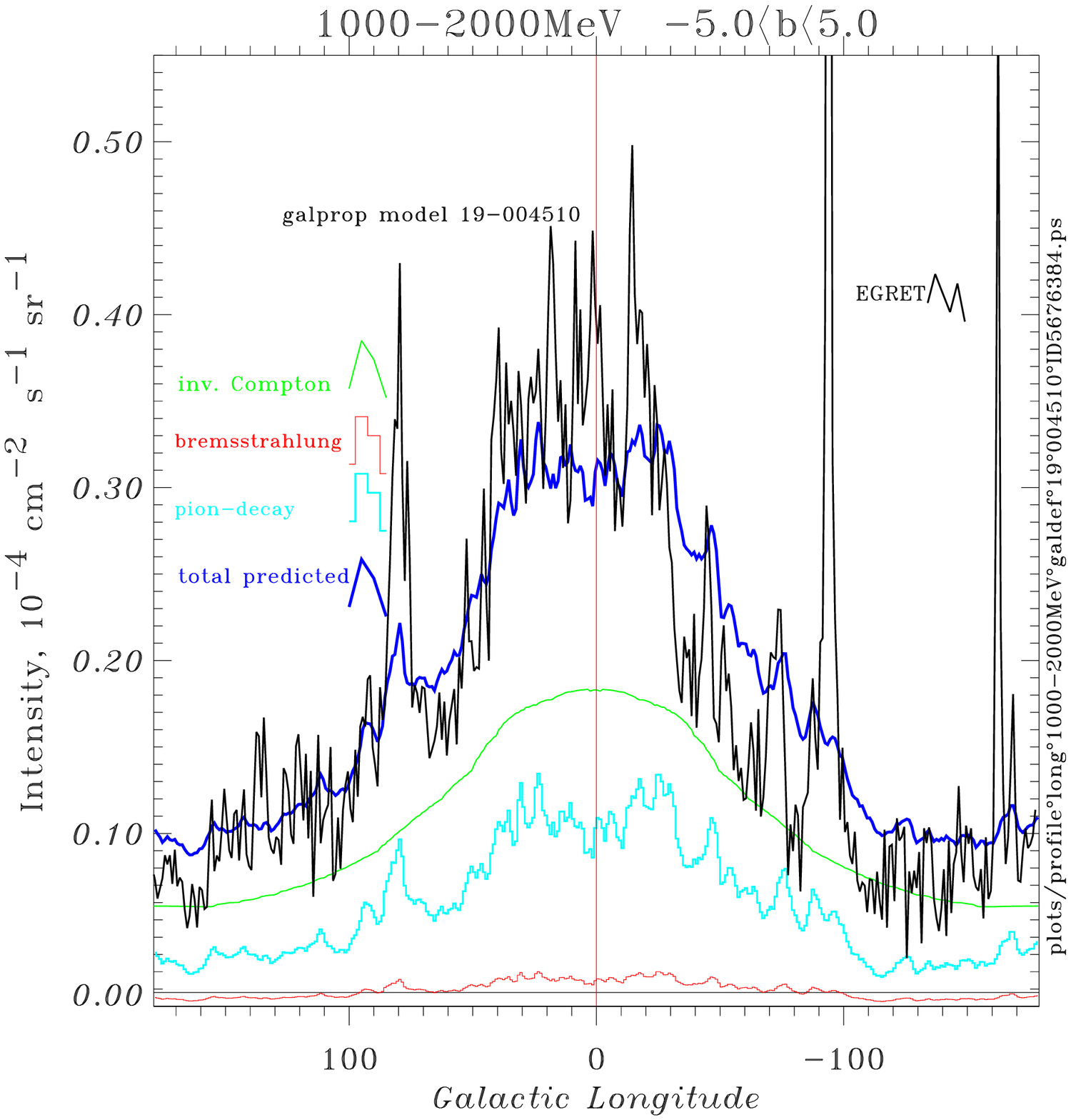,%
         height=\fhb,width=\fwb,clip=}}}
   \end{picture}
\caption{FIGURE 2. 
Left: Latitude distribution for 1--2 GeV (EGRET data),
compared to reacceleration model with hard electron spectrum.
Right: Longitude distribution for $|b|<5^\circ$.
Note that point sources have not been removed from the data.
\label{fig2} }
\end{figure}

\begin{figure}[!t]
   \begin{picture}(120,45)(5,-45)
      \put(-1,0){ \makebox(60,0)[tl]{ \psfig{file=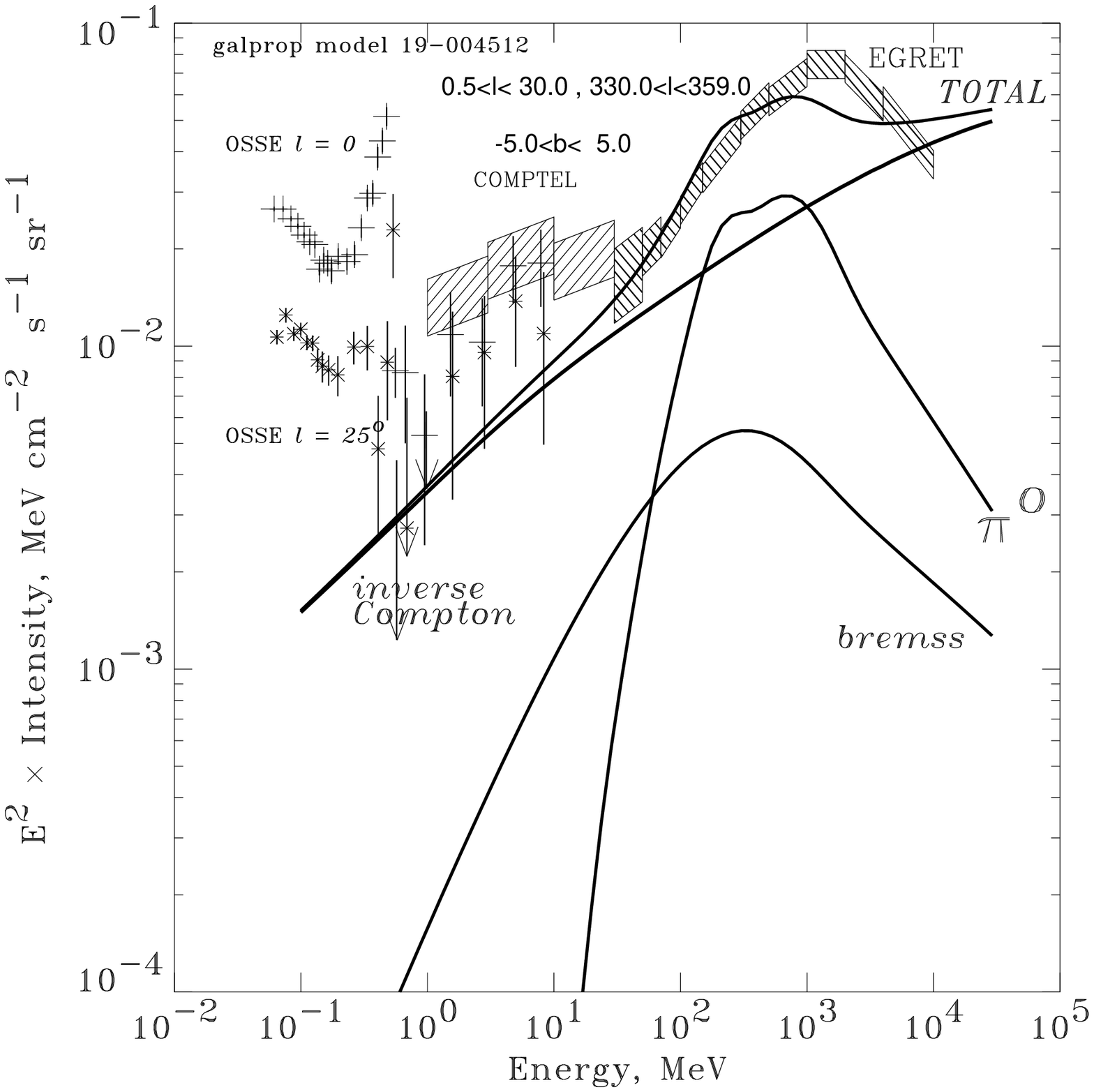,%
         height=\fhb,width=\fwb,clip=}}}
      \put(61,0){ \makebox(60,0)[tl]{ \psfig{file=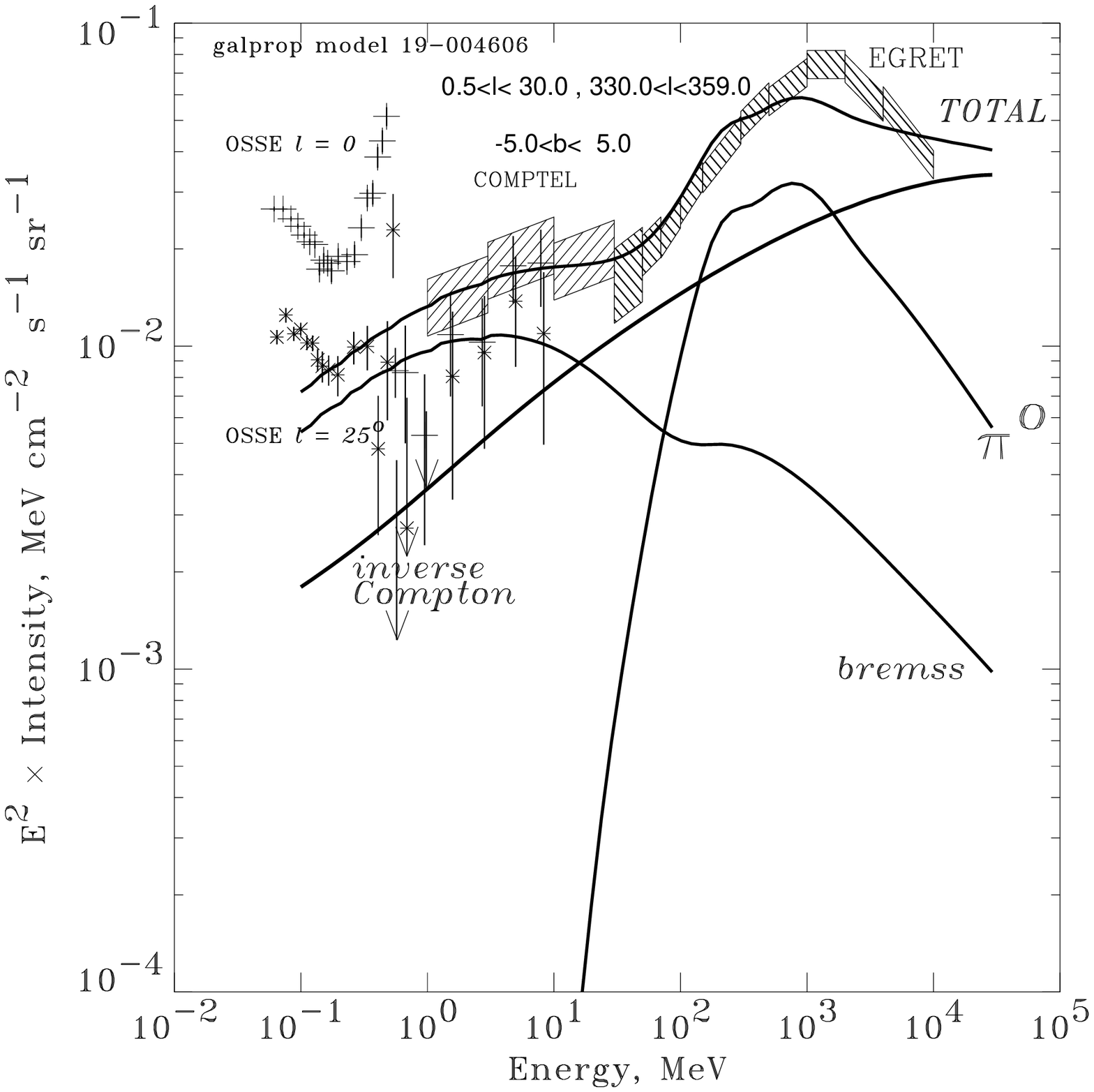,%
         height=\fhb,width=\fwb,clip=}}}
   \end{picture}
\caption{FIGURE 3. 
\gray spectrum of inner Galaxy (OSSE: Kinzer et al.\ 1997, COMPTEL:
Strong et al.\ 1998, EGRET: Strong \& Mattox 1996) compared to models
with a hard electron spectrum without (left) and with low-energy upturn
(right).
\label{fig3} }
\end{figure}

\bsk
\ni 2. MODELS 
\ssk
\ni 

We consider a propagation model with reacceleration using parameters
derived from isotopic composition (SM98).  A new calculation of the
ISRF has been made based on stellar population models and COBE
data. The electron injection spectral index is taken as --1.7 (with
reacceleration), which after propagation provides consistency with
radio synchrotron data.  Fig.~1 shows the electron spectrum at $R
= 8.5$ kpc in the disk for these models, and the synchrotron index.
Following PE98, for the present study we do not require consistency
with the locally measured electron spectrum above 10 GeV since the
rapid energy losses cause a clumpy distribution so that this is not
necessarily representative of the interstellar average.  The
$\pi^0$-decay $\gamma$-rays are calculated explicitly from the
propagated $p$ and He spectra (Dermer 1986, MS98).  A halo size
(distance from plane to boundary) of $z_h$=4 kpc is adopted, consistent
with our $^{10}$Be analysis (SM98).

\bsk
\ni
3. HARD ELECTRON SPECTRUM
\ssk
\ni

Fig.~2 shows the model latitude and longitude \gray distributions for
the inner Galaxy for 1--2 GeV, convolved with the EGRET point-spread
function, compared to Phase 1--4 data.  It shows that a model with
large IC component can indeed reproduce the data.  The latitude
distribution here is not as wide as at low energies owing to the rapid
energy losses of the electrons, so that an observational distinction
between a gas-related $\pi^0$-component from a hard nucleon spectrum
and the IC model does not seem possible on the basis of $\gamma$-rays
alone.  This model does fit above 100 MeV, but does not fit the \gray
spectrum below $\sim$30 MeV (Fig.~3 left).  In order to fit the
low-energy part as diffuse emission (Fig.~3 right) requires a rapid
upturn in the CR electron spectrum below 200 MeV (e.g., as in Fig.~1).
However, a population of unresolved sources seems more probable due to
the energetics problems (Skibo et al.\ 1997) and would be the natural
extension of the plane emission seen by OSSE and GINGA.

\bsk
\ni 4. TEST FOR A HARD NUCLEON SPECTRUM USING $\bar{p}$ AND $e^+$
\ssk
\ni 

Fig.~4 shows another possible origin for the $>$1 GeV excess, an
interstellar nucleon spectrum which is harder than observed locally
(MSR98).

The $\bar{p}/p$ ratio expected for this case and the `normal' spectrum
compared to recent data is shown in Fig.~5 (left) (MSR98).  Our
`normal' model calculation agrees with that of Simon et al.\ (1998).
For the case of a hard nucleon spectrum the ratio is consistent with
the data at low energies, but it is larger than the point at
3.7--19 GeV (Hof et al.\ 1996) by about $5\sigma$.  On the basis of the
$\bar{p}/p$ data point $\ge$3 GeV we seem already to be able to exclude
the hard nucleon spectrum, but confirmation of this conclusion must
await more accurate data at high energies.

Fig.~5 (right) shows the interstellar positron spectrum for these cases
(the formalism is given in MS98).  The flux for the `normal' case
agrees with recent data. For the hard nucleon spectrum the flux is
higher than observed; this provides more evidence against a hard
nucleon spectrum.  However this test is less direct than $\bar{p}$ due
to the difference in particle type and the large effect of energy
losses.

\begin{figure}[!t]
   \begin{picture}(120,43)(5,-45)
      \put(26,0){ \makebox(60,0)[tl]{ \psfig{file=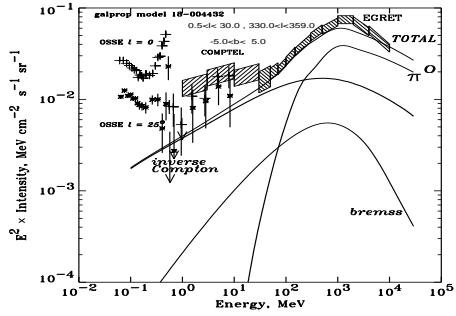,%
         height=\fhb,width=\fwb,clip=}}}
   \end{picture}
\caption{FIGURE 4. 
The same as in Fig.~3 but for a hard nucleon spectrum.
\label{fig4} }
\end{figure}
\begin{figure}[!t]
   \begin{picture}(120,43)(5,-45)
      \put(-1,0){ \makebox(60,0)[tl]{ \psfig{file=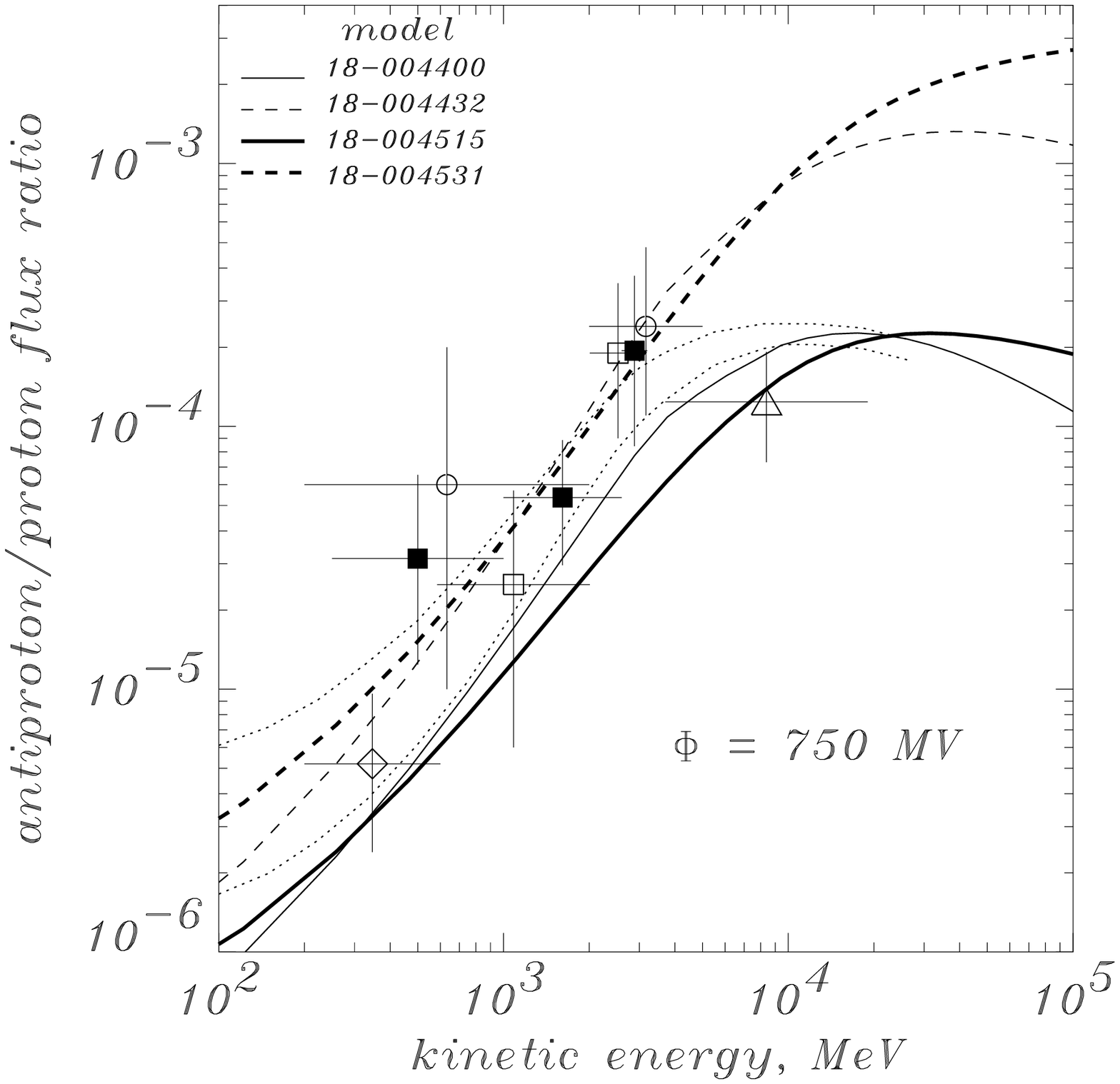,%
         height=\fhb,width=\fwb,clip=}}}
      \put(61,0){ \makebox(60,0)[tl]{ \psfig{file=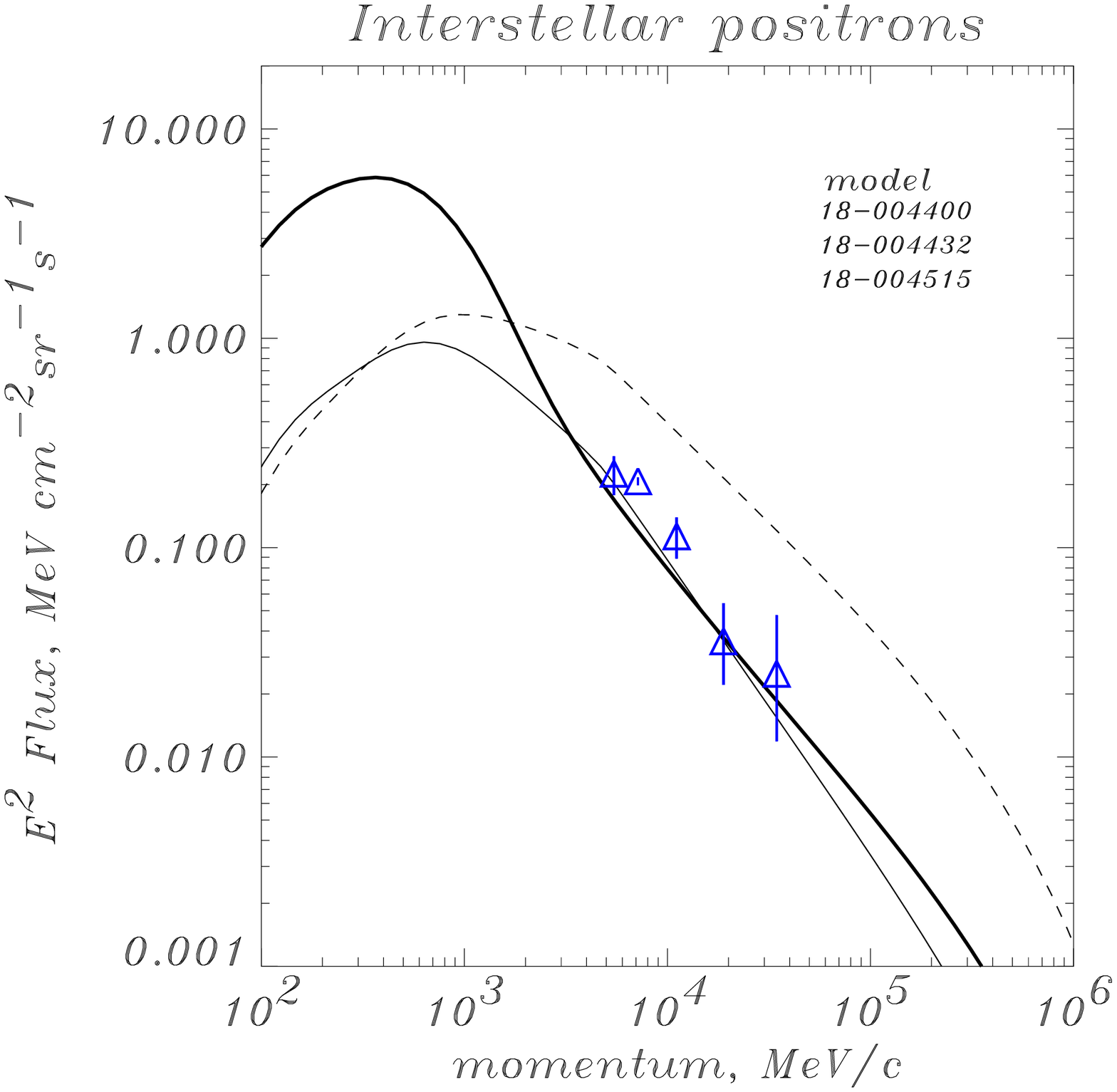,%
         height=\fhb,width=\fwb,clip=}}}
   \end{picture}

\caption{FIGURE 5. 
Left:  $\bar{p}/p$ ratio for the `normal' spectrum (solid
lines) and for the hard nucleon spectrum (dashes) used for the \gray
calculation. The thick lines show the case with
reacceleration.  Dotted lines: calculations of Simon et al.\ (1998).
Data: see references in MSR98.
Right: Spectra of secondary $e^+$'s for `normal' (thin line)
and hard (dashes) nucleon spectra (no reacceleration). 
Thick line: `normal' case with reacceleration.
Data: Barwick et al.\ (1998).}
\end{figure}

}

\bsk
\baselineskip = 12pt


{\references \ni REFERENCES
\ssk

\ref Barwick, S.W., et al.\ 1998, ApJ, 498, 779
\ref Dermer, C.D.\ 1986, A\&A, 157, 223
\ref Hof, M., et al.\ 1996, ApJ, 467, L33
\ref Kinzer, R.L., et al.\ 1997, ApJ, submitted (OSSE preprint \#89)
\ref Moskalenko, I.V., Strong, A.W.\ 1998, ApJ, 493, 694 (MS98)
\ref Moskalenko, I.V., Strong, A.W., Reimer, O.\ 1998, A\&A,
      338, L75 (MSR98)
\ref Pohl, M., Esposito, J.A.\ 1998, ApJ, 507, 327 (PE98)
\ref Simon, M., Molnar, A., Roesler, S.\ 1998, ApJ, 499, 250
\ref Skibo, J.G., et al.\ 1997, ApJ, 483, L95
\ref Strong, A.W., Mattox, J.R.\ 1996, A\&A, 308, L21
\ref Strong, A.W., Moskalenko, I.V.\ 1998, ApJ, 509, Dec.\ 10 issue (SM98)
\ref Strong, A.W., et al.\ 1997, in 4th Compton Symp., AIP 410,
      p.1198
\ref Strong, A.W., et al.\ 1998, in 3rd INTEGRAL Workshop,
      in press (astro-ph/9811211)
}                      

\end{document}